\documentclass[prl,letterpaper,twocolumn,showpacs]{revtex4}
\usepackage{times,xspace}
\usepackage{amsbsy,amssymb,amsmath,bm}
\usepackage{graphicx,color,epsfig,rotate}
\usepackage{fancyhdr}

\def\bbbc{{\mathchoice {\setbox0=\hbox{$\displaystyle\rm C$}\hbox{\hbox
to0pt{\kern0.4\wd0\vrule height0.9\ht0\hss}\box0}}
{\setbox0=\hbox{$\textstyle\rm C$}\hbox{\hbox
to0pt{\kern0.4\wd0\vrule height0.9\ht0\hss}\box0}}
{\setbox0=\hbox{$\scriptstyle\rm C$}\hbox{\hbox
to0pt{\kern0.4\wd0\vrule height0.9\ht0\hss}\box0}}
{\setbox0=\hbox{$\scriptscriptstyle\rm C$}\hbox{\hbox
to0pt{\kern0.4\wd0\vrule height0.9\ht0\hss}\box0}}}}

\newcommand{\ignore}[1]{}
\newcommand{\mComment}[1]{}
\newcommand{\gComment}[1]{}
\newcommand{\jComment}[1]{}
\newcommand{\rComment}[1]{}
\newcommand{\lComment}[1]{}

\renewcommand{\mComment}[1]{\textcolor{blue}{Manny: #1}}
\renewcommand{\gComment}[1]{\textcolor{red}{Gerardo: #1}}
\renewcommand{\jComment}[1]{\textcolor{green}{Jim: #1}}
\renewcommand{\rComment}[1]{\textcolor{magenta}{Ray: #1}}
\renewcommand{\lComment}[1]{\textcolor{purple}{Rolando: #1}}

\begin{document}
\title{Electronic Mechanism for the Coexistence of Ferroelectricity and Ferromagnetism}
\author{C. D. Batista$^1$, J. E. Gubernatis$^1$, and  Wei-Guo Yin$^2$}
\affiliation{$^1$Theoretical Division,
Los Alamos National Laboratory, Los Alamos, NM 87545 \\
$^2$Physics Department, Brookhaven National Laboratory, Upton, New York 11973-5000}

\date{Received \today }

\begin{abstract}
We study the strong coupling limit of a two-band Hubbard Hamiltonian
that also includes an inter-orbital on-site repulsive interaction
$U_{ab}$. When the two bands have opposite parity and are quarter filled, we prove that the
ground state is simultaneously ferromagnetic and ferroelectric for  infinite intra-orbital Coulomb
interactions $U_{aa}$ and $U_{bb}$. We also show that this coexistence leads 
to a singular magnetoelectric effect.
\end{abstract}


\maketitle %
\thispagestyle{fancy}

{\it Introduction.} The interplay between order parameters of
different nature opens the door for designing new multifunctional
devices whose properties can be manipulated with more than one
physical field. For instance, the spin and orbital electronic
degrees of freedom can order individually or simultaneously 
producing different phases. In particular,
orbital ordering can produce symmetry-breaking states like
orbital magnetism, ferroelectricity (FE), quadrupolar electric or
magnetic ordering, and other multipolar orderings. The
magnetoelectric multiferroics, such as $R$(Fe,Mn)O$_3$
\cite{Wang03,Kimura03,Lottermoser04,Aken04,Kimel05} and
$R$Mn$_2$O$_5$ \cite{Hur04,Hur04_prl}, are real examples of 
materials that combine distinct useful properties
within a single system. Recent studies of these
multiferroic materials have revived interest in the magnetoelectric
effect, i.e., the induction of polarization with an applied magnetic
field and magnetization with an applied electric field. To date,
materials that exhibit both ferromagnetism (FM) and FE are
rare \cite{Ederer04} because the transition metal ion of
typical perovskite ferroelectrics is in a nonmagnetic $d^0$
electronic configuration. Therefore, it is essential to explore
alternative routes to the coexistence of the FM and
FE \cite{Ederer04,Efremov04,Ederer05}.

Different mechanisms for FE involving electronic
degrees of freedom have been proposed. There are those in which
FE results from bond ordered states induced either by
electron-phonon coupling (Peierls instability) \cite{Littlewood79}
or by pure electron-electron Coulomb interactions \cite{Armando04}.
In these cases, the ferroelectric state is clearly nonmagnetic due
to the singlet nature of the covalent bonds. In contrast,
considering a system of interacting \emph{spinless} fermions with
two atomic orbitals of \emph{opposite} inversion symmetry (say the
$d$ and $f$ orbitals), Portengen {\it et al.} \cite{Portengen}
predicted that permanent electric dipoles are induced by spontaneous
$d-f$ hybridization when particle-hole pairs (excitons) 
undergo a Bose-Einstein condensation. This result was confirmed in
the strong coupling limit of an extended Falicov-Kimball spinless
fermion model where both bands are dispersive \cite{Batista02}. It
was also confirmed numerically in the intermediate coupling regime
by using a constrained path Monte Carlo approach \cite{Batista04}.

The critical question that emerges is: {\it How do
FE and magnetism interplay when real electrons,
instead of spinless fermions, are considered?}
In this Letter, we answer this question, proving that the  mechanism
proposed by Portengen {\it et al.} \cite{Portengen} can coexist with
magnetically ordered states. In this case, the single electron
occupying the effective (say $d-f$ hybridized) orbital
simultaneously provides an electric and a magnetic dipole moment, and
the Coulomb repulsion is sufficient to generate a
strong coupling between both of them.

We start from a two-band Hubbard Hamiltonian that includes an
inter-orbital on-site repulsive interaction $U_{ab}$. Like in the
spinless fermion case, this interaction provides the ``glue'' for the formation of 
excitons. At quarter filling and in the strong coupling limit, we map the low energy spectrum of
the two-band Hubbard model, $H$, into an effective spin-pseudospin
Hamiltonian, $H_{\rm eff}$, where the pseudospin represents the orbital degree of
freedom. We prove that in the limit of large intra-orbital
repulsive interactions $U_{aa},U_{bb} \rightarrow \infty$, $H_{\rm eff}$ has a
ferromagnetic ground state that can be partially or fully saturated. By
combining this result with the previous analysis for spinless
fermions \cite{Batista02,Batista04,Wei03}, we show that
FM and FE coexist, and 
a divergent magnetoelectric is demonstrated using the SO(4) symmetry of $H$. Our conclusions are reinforced by 
a semi-classical and a numerical computation of the zero temperature ($T=0$) 
phase diagram of $H$ that goes beyond the limiting case $U_{aa},U_{bb} \rightarrow \infty$.

{\it Hamiltonian.} We consider a two-band Hubbard model with a local
inter-band Coulomb interaction $U_{ab}$ on a $D$-dimensional
hypercubic lattice \cite{note}:
\begin{eqnarray}
H &=& \sum_{{\bf i},\eta,\nu,\nu',\sigma}
\!\! t_{\nu \nu'}(f^{\dagger}_{\bf i \nu \sigma} f^{\;}_{{\bf i+{\hat e}_{\eta}} \nu' \sigma}
+ f^{\dagger}_{{\bf i+{\hat e}_{\eta}} \nu' \sigma} f^{\;}_{\bf i \nu \sigma})
\nonumber \\
&+& \sum_{\bf i} U_{ab} n^a_{\bf i} n^b_{\bf i}
+ \sum_{\bf i, \nu} U_{\nu \nu} n^{\nu}_{\bf i \uparrow} n^{\nu}_{\bf i \downarrow}
+ \sum_{{\bf i},\nu} \epsilon_{\nu} n^{\nu}_{\bf i},
\end{eqnarray}
where $\eta=\{x,y,z,...\}$, $\nu=\{ a,b \}$, $n^{\nu}_{\bf i
\sigma}= f^{\dagger}_{\bf i \nu \sigma} f^{\;}_{\bf i \nu \sigma}$,
$n^{\nu}_{\bf i}= \sum_{\sigma} n^{\nu}_{\bf i \sigma}$ and $n_{\bf
i}= \sum_{\nu} n^{\nu}_{\bf i}$. Since the two orbitals, $a$ and
$b$, have opposite parity under spatial inversion, the inter-band
hybridization term must be odd under this operation:
$t_{ab}=-t_{ba}$. In addition, the intra-band hoppings $t_{aa}$ and
$t_{bb}$ will have {\it opposite} signs in general. The local spin
and pseudospin operators are given by the expressions:
\begin{equation}
s^{\mu}_{\bf i \nu}= \frac{1}{2} \sum_{\alpha \alpha'}
f^{\dagger}_{\bf i \nu \alpha} { \sigma}^{\mu}_{\alpha \alpha'}
f^{\;}_{\bf i \nu \alpha'},\;\;
\tau^{\mu}_{\bf i \sigma}=\frac{1}{2} \sum_{\nu \nu'}
f^{\dagger}_{\bf i \nu \sigma} { \sigma}^{\mu}_{\nu \nu'}
f^{\;}_{\bf i \nu' \sigma},
\end{equation}
where ${\sigma}^{\mu}$ are the Pauli matrices with $\mu=\{x,y,z\}$. The
total spin and pseudospin per site are: $s^{\mu}_{\bf i}=\sum_{\nu} s^{\mu}_{\bf i
\nu} $ and $\tau^{\mu}_{\bf i}=\sum_{\sigma} \tau^{\mu}_{\bf i \sigma} $.
The pseudospin component $\tau_x$ is proportional to the on-site hybridization.
Since the the orbitals $a$ and $b$ have opposite parity, the local electric dipole
moment is ${\bf p}_{\bf i} = {\boldsymbol \mu} \tau^x_{\bf i}$, where ${\boldsymbol \mu}$
is the dipole matrix element between the $a$ and $b$ orbitals \cite{Batista02,Batista04}.

Symmetry is a useful concept for describing the coexistence of different order parameters.\cite{Ortiz04} For $t_{aa}=\pm t_{bb}$, $t_{ab}=t_{ba}=0$, $\epsilon_a - U_{aa}/2= \epsilon_b - U_{bb}/2$, and $U_{aa}=U_{ab}=U_{bb}$, $H$ is invariant under a
U(1)$\otimes$SU(4) symmetry group. The U(1) symmetry corresponds to the conservation
of the total number of particles. The generators of the SU(4) symmetry group are the three components of the 
total spin $s^{\mu}_T= \sum_{\bf i} s^{\mu}_{\bf i}$ and pseudospin $\tau^{\mu}_T = \sum_{\bf i} {\tau}^{\mu}_{\bf i}$ plus
the nine operators:
\begin{equation}
\pi^{\mu \mu'}=\frac{1}{2} \sum_{{\bf i},\alpha \nu \alpha' \nu'}
f^{\dagger}_{{\bf i} \nu \alpha} {\sigma}^{\mu}_{\nu \nu'}
{\sigma}^{\mu}_{\alpha \alpha'} f^{\;}_{{\bf i} \nu' \alpha'}.
\end{equation}
The total spin is conserved for any set of parameters. 
If we just impose the condition $t_{ab}=t_{ba}=0$,  the
symmetry group of $H$ is reduced to the the subgroup
U(1)$\times$U(1)$\times$SO(4). The six generators of the SO(4) group
are the three components of total spin $s^{\mu}_T$ and the three
operators $\pi^{z \mu}$. This symmetry arises from separate total
spin and charge conservation of each band, as the two
bands are only coupled by the Coulomb interaction $U_{ab}$. The
symmetry operators $s^{\mu}_T \pm \pi^{z \mu}$ are the generators of
global spin rotations on each individual band (with the
$+$ sign for the $a$ band and the $-$ sign for the $b$ band).

{\it Strong Coupling Limit.} We will consider from now on  the quarter filled case $n_a+n_b=1$,
where $n_a$ and $n_b$ are the particle densities of the bands $a$ and $b$.
When $U_{aa},U_{bb},U_{ab} \gg |t_{\nu\nu'}|$, the low energy spectrum of $H$ can be mapped
to an effective spin-pseudospin Hamiltonian by means of a canonical transformation that eliminates the linear
terms in $t_{\nu\nu'}$:
\begin{eqnarray}
&H&_{\!\!\! \!\!\rm eff}= \! \sum_{{\bf i},\eta} [ \sum_{\mu} J_{\mu} {\tau^{\mu}_{\bf i}}  {\tau^{\mu}_{{\bf i} + {\bf e}_{\eta}}}
+ J_{xz} ({\tau^{z}_{\bf i}}  {\tau^{x}_{{\bf i} + {\bf e}_{\eta}}} - {\tau^{x}_{\bf i}}  {\tau^{z}_{{\bf i} + {\bf e}_{\eta}}})]
H^{H}_{{\bf i},\eta}
\nonumber \\
&+& \!\!\! \sum_{{\bf i},\eta}
{[}\frac{J_0}{2}+J_{1}(\tau^{z}_{\bf i} + \tau^{z}_{{\bf i} + {\bf e}_{\eta}})-J_2 (\tau^{x}_{\bf i} -
\tau^{x}_{{\bf i} + {\bf e}_{\eta}}){]}
(H^{H}_{{\bf i},\eta} - \frac{1}{2})
\nonumber \\
&+&  \!\!\! \sum_{{\bf i},\eta} [J'_z \tau^{z}_{\bf i}
\tau^{z}_{{\bf i} + {\bf e}_{\eta}}+ J'_{xz} ({\tau^{z}_{\bf i}}
{\tau^{x}_{{\bf i} + {\bf e}_{\eta}}} - {\tau^{x}_{\bf i}}
{\tau^{z}_{{\bf i} + {\bf e}_{\eta}}})]+ \sum_{\bf i} B_z
\tau^{z}_{\bf i}, \nonumber \label{KK}
\end{eqnarray}
where $H^{H}_{{\bf i},\eta}={\bf s_i}\cdot {\bf s_{{\bf i} + {\bf e}_{\eta}}} + \frac{1}{4}$
and
\begin{eqnarray}
J_z&=&\frac{4(t_{bb}^2-t_{ab}^2)}{U_{bb}}+\frac{4(t_{aa}^2-t_{ab}^2)}{U_{aa}},
\nonumber \\
J_{x}&=&\frac{8(t_{bb} t_{aa} - t_{ab}^2)}{U_{ab}}, J_{y}=\frac{8(t_{bb} t_{aa} + t_{ab}^2)}{U_{ab}},
\nonumber \\
J_{xz} &=& 4t_{ab} [\frac{(t_{aa}+t_{bb})}{U_{ab}}-\frac{t_{aa}}{U_{aa}}-\frac{t_{bb}}{U_{bb}}]
\nonumber \\
J_0 &=& \frac{2(t_{bb}^2+t_{ab}^2)}{U_{bb}}+\frac{2(t_{aa}^2+t_{ab}^2)}{U_{aa}},
J_1 = \frac{2t_{bb}^2}{U_{bb}}-\frac{2t_{aa}^2}{U_{aa}},
\nonumber \\
J_2 &=& 2 t_{ab} [\frac{(t_{aa}-t_{bb})}{U_{ab}}+\frac{t_{aa}}{U_{aa}}-\frac{t_{bb}}{U_{bb}}]
\nonumber \\
J'_z &=& \frac{2(t_{aa}^2+t_{bb}^2-2t_{ab}^2)}{U_{ab}}-\frac{J_z}{2},
\nonumber \\
J'_{xz} &=& 2t_{ab}
[\frac{(t_{aa}+t_{bb})}{U_{ab}}+\frac{t_{aa}}{U_{aa}}+\frac{t_{bb}}{U_{bb}}],
\label{pert}
\end{eqnarray}
and $B_z=\epsilon_a - \epsilon_b +(U_{bb}-U_{aa})/2$.
In this limit, because the double-occupancy is
forbidden in the low energy Hilbert space of $H_{\rm eff}$,
both ${\bf s}_{\bf i}$ and  ${\bf \tau}_{\bf i}$ belong to the $S=1/2$ representation
of the su(2) algebra. The
first  two terms of $H_{\rm eff}$ couple the spin and
the orbital degrees of freedom. As usual, the Heisenberg
antiferromagnetic interaction,  $H^{H}_{{\bf i},\eta}$,
is a direct consequence of the Pauli exclusion principle. On the
other hand, the anisotropic Heisenberg-like
pseudospin-pseudospin interaction reflects the
competition between an excitonic crystallization or staggered orbital ordering (SOO)
induced by the Ising term, and a Bose-Einstein condensation of
excitons induced by the $XY$-term \cite{Batista02,Wei03}. The first term of
$H_{\rm eff}$ shows explicitly that the amplitude of the excitonic
kinetic energy (or $XY$-pseudospin) term gets maximized
when the excitons are in a fully polarized ferromagnetic spin state.
However, antiferromagnetism (AF) is clearly favored by the second term.

{\it Large $U_{aa}$, $U_{bb}$ limit.} We will first prove that there are
partially and fully polarized ferromagnetic ground states of $H_{\rm eff}$ in
the limit of $U_{aa}$, $U_{bb} \rightarrow \infty$ and $t_{ab}=0$, and that
the total spin or magnetization $s_T$ take the values $\tau^z_T \leq s_T \leq N/2$. After proving this result, we will show that these ferromagnetic solutions are also 
ferroelectric for $B^{c1}_z \leq B_z < B^{c2}_z$ and, using the SO(4) symmetry,
we will derive an exact expression for ground state electric polarization as a 
function of the magnetization.

Since $J_0$ and $J_1$ vanish in this limit, $H_{\rm eff}$ is reduced
to:
\begin{eqnarray}
{\bar H}_{\rm eff} &=& \sum_{\langle \bf i,j \rangle} [{J'_z} {\tau^z_i}  {\tau^z_j}
+ J_{\perp}
(\tau^{x}_{\bf i} \tau^{x}_{\bf j}+\tau^{y}_{\bf i} \tau^{y}_{\bf j})
H^{H}_{\bf i,j}]
+  B_z \sum_{\bf i} \tau^z_{\bf i}, \nonumber
\end{eqnarray}
where the angular brackets indicate that the sum is over nearest-neighbors,
$J_{\perp}=J_x=J_y$, and $H^{H}_{\bf i,j}={\bf s_i}\cdot {\bf s_j} + \frac{1}{4}$.
To prove our statement we will use a basis of eigenstates of the local operators $\tau^z_{\bf i}$ and $s^z_{\bf i}$:
$\{ | \tau^z_1 ... \tau^z_N \rangle \otimes | s^z_1 ... s^z_N\rangle \}$ where $N$ is the total number of sites.
The off-diagonal matrix elements of ${\bar H}_{\rm eff}$
\begin{eqnarray}
\langle s'^z_N ...s'^z_1| \otimes \langle \tau'^z_N ... \tau'^z_1 |  {\bar H}_{\rm eff}
| \tau^z_1 ... \tau^z_N \rangle \otimes | s^z_1 ... s^z_N\rangle =
\nonumber \\
2 J_{\perp} \sum_{\langle \bf i,j \rangle} \langle \tau'^z_N ... \tau'^z_1 |
 H^{xy}_{\bf i,j}| \tau^z_1 ... \tau^z_N \rangle
\langle s'^z_N ...s'^z_1| H^{H}_{\bf i,j}| s^z_1 ... s^z_N\rangle
\end{eqnarray}
are non-positive because $J_{\perp}<0$, and the matrix elements of
$H^{xy}_{\bf i,j}=\tau^{x}_{\bf i} \tau^{x}_{\bf j}+\tau^{y}_{\bf i}
\tau^{y}_{\bf j}$ and $H^{H}_{\bf i,j}$ are explicitly
non-negative. According to the generalized Perron's theorem (see for
instance Ref.~\cite{Horn}), there is one ground state of ${\bar
H}_{\rm eff}$,
\begin{equation}
|\Psi^g \rangle = \sum_{\{\tau^z \}}  | \tau^z_1 ... \tau^z_N \rangle \otimes
\sum_{\{s^z, s^z_T=0 \}} a_{\{\tau^z\},\{s^z\}} | s^z_1 ... s^z_N\rangle,
\end{equation}
such that all the amplitudes $a_{\{\tau^z\},\{s^z\}}$ are non-negative ( $\{\tau^z\}$ and
$\{s^z \}$ denote all the possible configurations of $\tau^z_{\bf i}$ and $s^z_{\bf i}$).
We can rewrite $|\Psi^g_{s^z_T=0} \rangle$ in the following way:
\begin{equation}
|\Psi^g \rangle = \sum_{\{\tau^z \}'}  b_{\{\tau^z\}} | \tau^z_1 ... \tau^z_N \rangle \otimes
|\Phi \{\tau^z \}\rangle,
\end{equation}
with
\begin{eqnarray}
|\Phi \{\tau^z \} \rangle &=& \frac{1}{b_{\{\tau^z\}}} \sum_{\{s^z, s^z_T=0 \}} a_{\{\tau^z\},\{s^z\}} | s^z_1 ... s^z_N\rangle,
\nonumber \\
{b_{\{\tau^z\}}} &=& \sqrt{\sum_{\{s^z, s^z_T=0 \}}
a^2_{\{\tau^z\},\{s^z\}}}.
\end{eqnarray}
The set ${\{\tau^z \}'}$ corresponds to all the configurations of the $\tau^z_{\bf i}$ variables such that
${b_{\{\tau^z\}}}>0$. Note that each spin state $|\Phi \{\tau^z \} \rangle$ is normalized and
$\sum_{\bf i} b^2_{\{\tau^z\}}=1$ because $|\Psi^g \rangle$ is also normalized. The ground state energy
of ${\bar H}_{\rm eff}$ is :
\begin{eqnarray}
\langle \Psi^g |{\bar H}_{\rm eff} |\Psi^g \rangle = \langle \Psi^g |\sum_{\langle \bf i,j \rangle} {J_z} {\tau^z_i}  {\tau^z_j}
+  B_z \sum_{\bf i} \tau^z_{\bf i}|\Psi^g \rangle +
\nonumber \\
2 J_{\perp} \sum_{\langle \bf i,j \rangle} \sum_{\{\tau^z,\tau'^z  \}'}  b_{\{\tau^z\}} b_{\{\tau'^z\}} \langle \tau'^z_N ... \tau'^z_1 |
 H^{xy}_{\bf i,j}| \tau^z_1 ... \tau^z_N \rangle \times
\nonumber \\
\langle \Phi \{\tau'^z \}| H^{H}_{\bf i,j}| \Phi \{\tau^z \}
\rangle. \label{gse}
\end{eqnarray}
$H^{H}_{\bf i,j}$ is a bounded operator with eigenvalues $\pm 1/2$. Therefore
$\langle \Phi \{\tau'^z \}| H^{H}_{\bf i,j}| \Phi \{\tau^z \} \rangle \leq 1/2$ and the equality holds
for any pair of nearest-neighbor sites $({\bf i, j})$ if $| \Phi \{\tau^z \} \rangle = |s^z_T, s_T=N/2 \rangle$ where
$|s^z_T, s_T=N/2 \rangle$ is the state with maximum total spin (fully polarized) and $\sum_{\bf i} s^z_{\bf i}=s^z_T$.
Hence, expression (\ref{gse}) is minimized for the fully polarized spin configuration which means that there is family of
ground states $|\Psi^g \rangle$ that have maximum total spin $s_T$ and different values of $s^z_T$:
\begin{equation}
|\Psi^g \rangle = \sum_{\{\tau^z \}'}  b_{\{\tau^z\}} | \tau^z_1 ...
\tau^z_N \rangle \otimes |s^z_T, s_T=N/2 \rangle.
\end{equation}
This proves that there is a fully polarized ferromagnetic ground state 
of ${\bar H}_{\rm eff}$. Note the 
proof is valid in any dimension and is also valid for non-bipartite
lattices. When we restrict ${\bar H}_{\rm eff}$ to the subspace with
maximum total spin ($s_T=N/2$), the operator $({\bf S_i}\cdot {\bf
S_j} + \frac{1}{4})$ is replaced by $1/2$ and the restricted
Hamiltonian ${\bar H}^{FM}_{\rm eff}$ becomes exactly the same as
the one obtained in Ref.~\cite{Batista02} for the strong coupling
limit of a spinless extended Falicov-Kimball model \cite{Falicov}.
As shown in Ref.~\cite{Batista02,Wei03}, the quantum phase diagram
of ${\bar H}^{FM}_{\rm eff}$ contains a ferroelectric phase for
$B^{c1}_z \leq B_z < B^{c2}_z$, i.e., for a non-zero value of
$\tau^z_T$.

The SO(4) symmetry of ${\bar H}_{\rm eff}$ implies that the ground state
degeneracy is higher than  the $2s_T+1=N+1$ multiplet obtained from
the global SU(2) spin rotations. Ground states with different total
spin can also be obtained by making different global spin rotations
for the bands $a$ and $b$.  The spins of each individual band will
remain fully polarized under these transformations, but the relative
orientation between spins of different bands will change. In
particular, the minimum total spin will occur when spins in
different bands are ``anti-aligned'', i.e., $s_T=\tau^z_T$. This
implies that the total spin of the ground state can take the values:
$\tau^z_T \leq s_T \leq N/2$. 

\begin{figure}[htb]
\vspace*{-0.9cm}
\includegraphics[angle=-90,width=8cm]{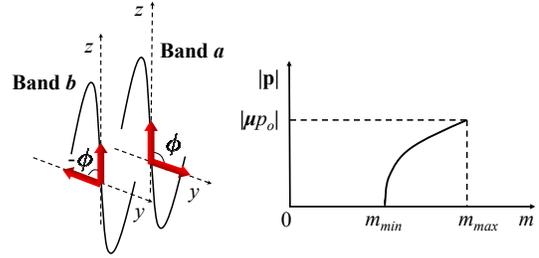}
\vspace*{-1.4cm}
\label{figu1}
\caption{Evolution of the magnetization (arrows) on each band under the SO(4) transformation 
$U(\phi)$. The ${\bf P}$ vs ${\bf m}$ plot shows the change of the 
electric dipole moment when the total magnetization evolves from the minimum value,  
obtained for $\phi=\pi/2$, to the maximum value $m(\phi=0)= 1/2$.}
\end{figure} 

To derive the magneto-electric effect, we define $| \chi \rangle$ as the fully polarized ferromagnetic 
and ferroelectric state obtained for $B^{c1}_z \leq B_z < B^{c2}_z$. Then, the average magnetization and 
electric dipole moment per site, $m$ and ${\bf p}$, are given by:
\begin{equation}
m=\frac{\sqrt{\langle \chi | {\bf s}^2_T | \chi \rangle}}{N} = \frac{1}{2} \sqrt{1+\frac{2}{N}}, \;\;\;
{\bf p}={\boldsymbol \mu} p_0 ,
\end{equation}
where $p_o = \langle \chi | \tau^x_T | \chi \rangle/N$.
A given ground state in the SO(4) multiplet can expressed as 
$|g_u\rangle= U | \chi \rangle$, where $U$ is an element of the SO(4) group.
In particular, choosing the set of transformations $U(\phi)= e^{i \phi \sum_{\bf j} (S^x_{{\bf j}a} - S^x_{{\bf j}b})}$
we get for ${\bf p}(\phi)={\boldsymbol \mu}\langle g_{u(\phi)}| \tau^x_T  |g_{u(\phi)} \rangle/N$ and 
$m(\phi)=\sqrt{\langle g_{u(\phi)}| {\bf s}^2_T |g_{u(\phi)} \rangle}/N$:
\begin{eqnarray}
m(\phi)= &=&
\frac{1}{2} \sqrt{1-4 n_a n_b \sin^2{\!\!\phi} +\frac{2}{N}},
\nonumber \\
{\bf p}(\phi) &=& {\boldsymbol \mu} p_o \cos{\phi},
\label{pola}
\end{eqnarray}  
where we have used that ${\bf s}_T= {\bf s}_{aT} + {\bf s}_{bT}$ and that $U(\phi)$ rotates the vectors 
${\bf s}_{aT}$ and ${\bf s}_{bT}$ around the $x$-axis by angles $\phi$ and $-\phi$ respectively. 
For the second relation, we have used that:
\begin{equation}
U^{\dagger}(\phi) \tau^x_T U^{\;}(\phi)= \cos{\phi} \; \tau^x_T - \sin{\phi} \; \pi^{yx},
\end{equation}
and $\langle \chi | \pi^{yx} | \chi \rangle = 0$. For $\phi=\pi/2$,
the magnetization of the $a$ and $b$ bands have opposite sign (see Fig.1) and 
the total magnetization per site is minimized: $m_{min}=m(\pi/2)=|n_a-n_b|/2$ (we have taken
the thermodynamic limit $N\rightarrow \infty$). The electric polarization can be expressed as 
a function of $m$ by combining Eqs.~(\ref{pola}):
\begin{equation}
|{\bf p}|=2 |{\boldsymbol \mu}p_0| \frac{\sqrt{m^2-m^2_{min}}}{\sqrt{1-4m_{min}^2}}.
\label{pvsm}
\end{equation}
The electric dipole moment is zero for $m=m_{min}$ and it increases as $\sqrt{m-m_{min}}$
implying that the derivative $d|P|/dm$ diverges at $m=m_{min}$ as $1/\sqrt{m-m_{min}}$.
This important result shows that the interplay between the spin and the orbital degrees of freedom can produce an 
{\it enormous magneto-electric effect} (see Fig.~1). 
\begin{figure}[htb]
\vspace*{-0.3cm}
\includegraphics[angle=270,scale=0.4]{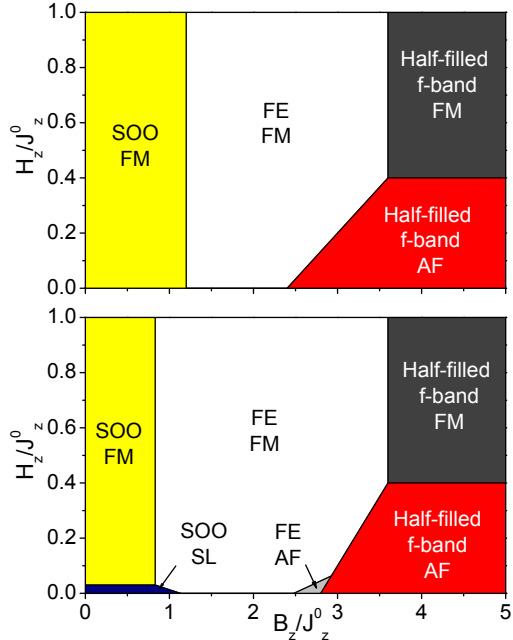}
\vspace*{-0.5cm}
\label{figu2}
\caption{Zero temperature phase diagram of the two dimensional version of $H_{\rm eff}$ 
plus a Zeeman term $H_z s^z_T$ computed in the spin-wave approximation (top) and by exact diagonalization of a $4\times4$ cluster (bottom). The parameter values are $J_0= 0.2 J_z^0$, $J_x=J_y=-1.6 J_z^0$, $J_1=-0.1 J_z^0$
and $J_z^0=J'_z+J_z/2$.}
\end{figure} 

It is important to verify the stability of the coexisting ferroelectric and ferromagnetic
phases away from the limit we have considered above. For this purpose, we computed  
the $T=0$ diagram of the two dimensional version of $H_{\rm eff}$ plus a Zeeman term, $H_z s^z_T$,
in a spin-wave approximation (top of Fig.~2) and by  exact diagonalization of a $4\times4$ 
cluster (bottom of Fig.~2). In this case, the values of of $U_{aa}$ and $U_{bb}$ are finite ($J_0= 0.2 J_z^0$), 
and the coexistence of FM and FE obtained for $H_z=0$ and $B^{c1}_z \leq B_z < B^{c2}_z$ 
is still present (see Fig.~2). For $B_z < B^{c1}_z$ and $H_z=0$, 
the spin-wave phase diagram exhibits coexistence of FM and SOO
\cite{Batista02}. The FM ordering is replaced by a spin liquid (short ranged spin-spin correlations) in the  
phase diagram computed by exact diagonalization indicating that quantum fluctuations play a major role in that regime of 
parameters. As expected, for large enough $B_z $ the system becomes an antiferromagnetic 
Mott insulator (one band is a half-filled and the other one is empty). The exact diagonalization 
shows again that quantum fluctuations generate an intermediate phase between the FE-FM state 
and the Mott insulator in which FE and AFM coexist (see bottom of Fig.~2).
For high values of $H_z$, the system is fully polarized and the both $T=0$ phase diagrams 
coincide with the one obtained in Ref.~\cite{Batista02}.

The ferromagnetic state
can be further stabilized by the inclusion of the
\emph{ferromagnetic} on-site inter-orbital exchange
interaction. For example, the intra-atomic $4f-5d$ exchange
interaction is about 0.2 eV in EuB$_6$ \cite{Kunes,Li}.
The other important aspect to consider is the role of a finite
inter-band hybridization $t_{ab}$. Exact diagonalization results
\cite{Wei} show that the lowest total spin ground state,
$s_T=\tau^z_T$ or $m=m_{min}$,  is the one stabilized after the inclusion of a
small  $t_{ab}$ term. According to Eq.(\ref{pvsm}), this unsaturated ferromagntic state 
gives rise to a divergent magnetoelectric effect. In this situation, a small increase in
the magnetization produced by an applied magnetic field will generate a { \it large increase of
the  electric dipole moment} in the way depicted in Fig.~1.

In summary, we have shown that the electron-electron Coulomb
interaction can produce coexisting FM and
FE. Both phases arise simultaneously from the
condensation of excitons or particle-hole pairs that exist in two
bands with opposite parity under spatial inversion. The coexistence
requires the presence of large intra-orbital Coulomb interactions to
reduce the strength of the antiferromagnetic interaction. We have 
also shown that the  coexistence of FE and
FM leads to a divergent magnetoelectric effect.
In the proximity of the ferroelectric-ferromagnetic instability, a
small magnetic field can produce an enormous change in the electric
polarization.


We thank W. E. Pickett and J. Kune\v{s} for pointing out
Ref.~\cite{Li}.  LANL is
supported by US DOE under Contract No. W-7405-ENG-36.
BNL is supported by US DOE under Contract
No. DE-AC02-98CH1-886.


\begin{references}
\bibitem{note}
This is the minimal Hamiltonian for studying the interplay between electronic 
ferroelectricity and magnetism. In real systems, the number of orbitals that lie
close to the Fermi level after the crystal filed splitting can be higher than two.
\end{references}
\end{document}